# Inversely Prepolarized Piezoceramic Cantilever.

D.P. Sedorook and I.V. Ostrovskii

*Department of Physics and Astronomy, University of Mississippi, Oxford, MS 38677, USA*

A nonuniform piezoelectric cantilever with enhanced amplitude of vibration at resonances is proposed. The cantilever made of lead zirconate titanate (type of PZT-5H) containing inversely poled piezoelectric domains shows an advantageous increase in vibration amplitude when compared to a single domain device. The amplitude of vibrations is enhanced when the domain boundaries are located at the nodes of vibration displacement. The vibration amplitude of a cantilever with 2 or 3 periodically inverted domains is mostly affected at the 2$^{nd}$ or 3$^{rd}$ resonance frequency, respectively. The amplitudes are calculated using the Finite Element method (FEM).

Piezoelectric cantilevers are the key components of microelectromechanical systems with various applications[1], which include atomic force microscopy[2], detection of chemical compounds[3], and the actuation of wing flapping mechanisms in micromechanical robotic insects[4,5]. In micromechanical insect prototype applications, rectangular cantilever dimensions are oblong in shape and transverse vibration is utilized[5]. Cantilevers that are less oblong in shape may also be good microactuators when the longitudinal displacement is utilized[6]. The piezoelectric material, polyvinylidene fluoride (PVDF), has been used to demonstrate this effect to some extent by designing a system with a three layer PVDF-shim metal composite laminate, which lead to a PVDF modal actuator prototype[7]. Also implemented were two inversely poled PVDF-films situated side-by-side on the shim metal along the structure width[8]. In reference [8], a bimorph PVDF actuator was proposed and calculated by using the Finite Element Method (FEM). The FEM was also used to calculate proper electrode patterns of a three layer PVDF laminate composite plate[9]. We have to note that in the cited publications [7,8,9], a low frequency region of hundreds of Hertz was considered and used in the experiments. However, the actuation effect that has been established could be much further enhanced if a material with a higher piezoelectric coupling coefficient than PVDF is used, for example PZT. Despite it is a bulk material, nowadays so-called nano-ceramic can be found on the market place, which opens new perspectives for modern applications.

Different types of PZT ceramic are commonly used in cantilevers due to the high piezoelectric coupling of this material. A cantilever normally utilizes the zeroth order vibration mode $A_0$ with the dominant displacement in the direction perpendicular to the piezoelectric plate. In a plate that is less oblong, the longitudinal mode $S_0$ may cause the total displacement to be in the longitudinal direction.

In this paper, we show that the longitudinal pulsation can be reduced to the advantage of the transverse vibration by introducing a set of inversely poled piezoelectric domains along the length of the plate. A single crystal cantilever and a 2-domain cantilever with 2 inversely poled domains are shown in Figure 1(a) and (b), respectively. The Z axis of the PZT-5H material PIC-181 is perpendicular to the cantilever plate, which vibrates in the ZX plane. Both cantilevers have dimensions 22.4×0.7 mm along the X and Z directions respectively. The white block arrows in Figure 1 represent the polarization of piezoelectric domains. The length of the larger domain in Figure 1(b) is 16.1 mm. The smaller domain is 4.9 mm-long. The interdomain boundary is at $x = 17.5$ mm because the vertical component of acoustical displacement $u_z$ is equal to zero at this point for the second resonance of the single domain and 2-domain cantilevers. Two metal electrodes are deposited on the top ($z = 0.7$ mm) and the bottom ($z = 0$) surfaces. Normally, the bottom electrode is grounded, while an AC voltage up to 1 kV is applied to the top electrode. The resulting electric field inside the cantilever is well below the coercive field for piezoelectric ceramic [10]. The device is designed to operate at a temperature that is lower than the Curie temperature to prevent depolarization of the domains. The multidomain cantilevers with unipolar electrode design having one, two, and three inversely pooled ferroelectric domains are presented in Fig. 1.

(1)

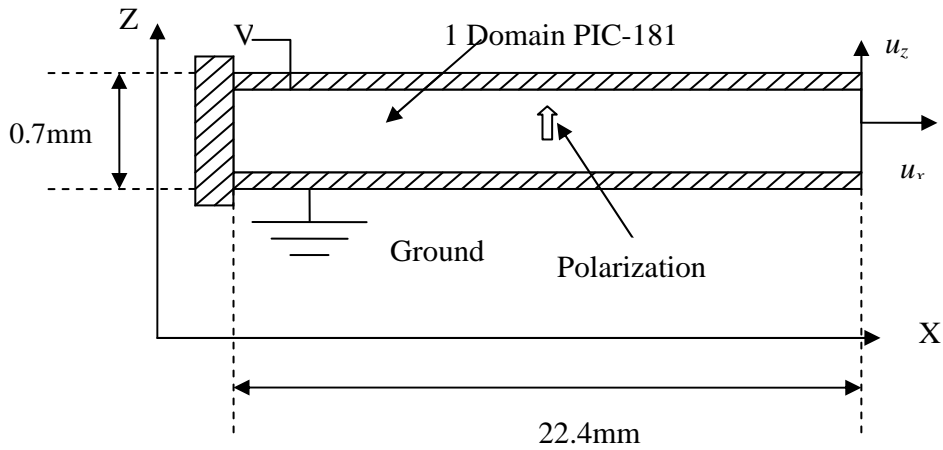

(2)

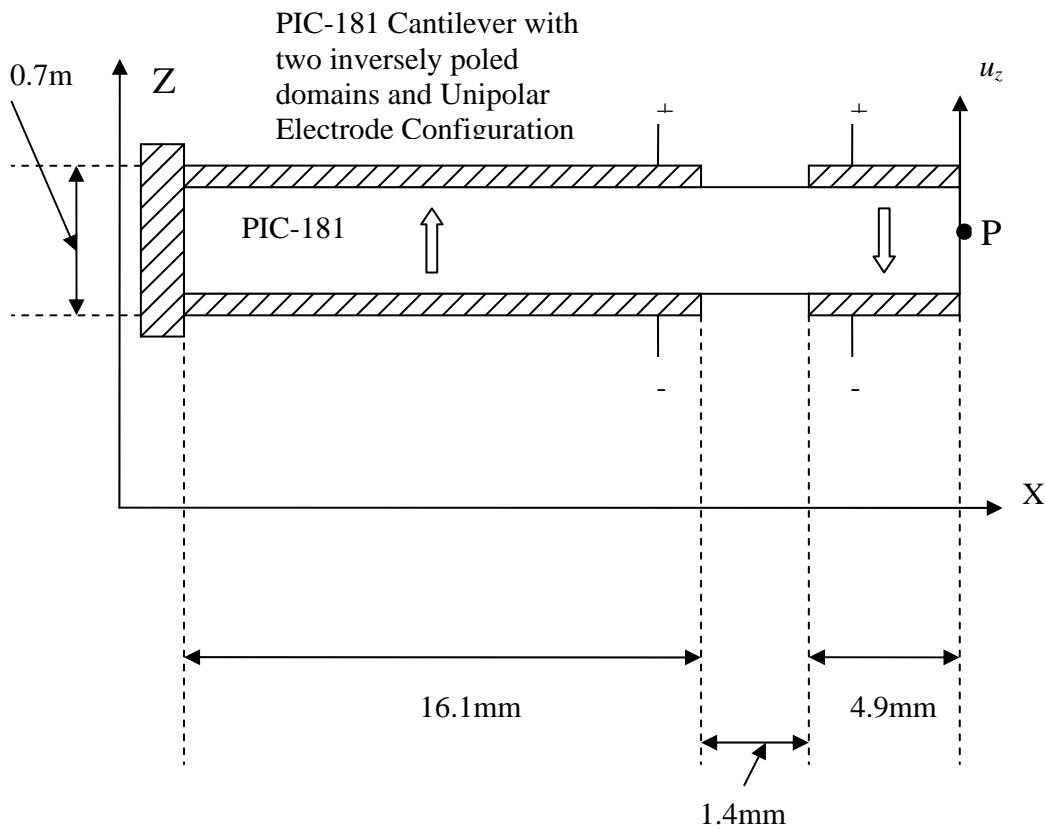

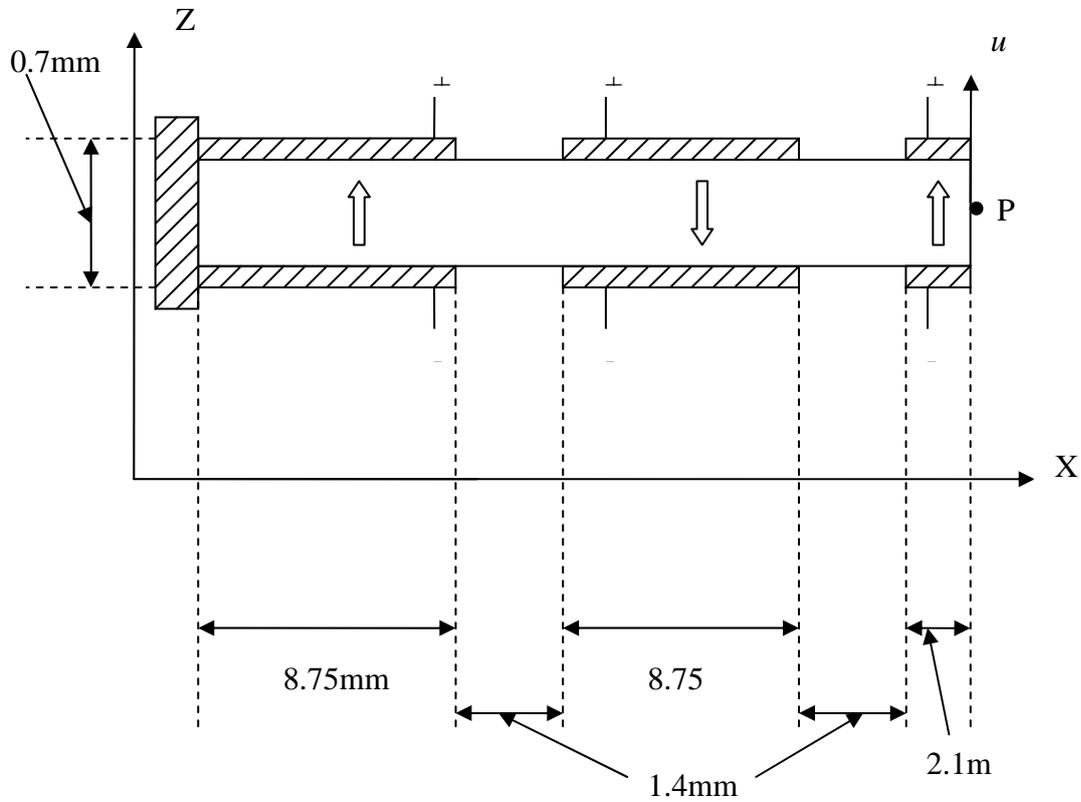

Figure 1. The multidomain cantilevers with the unipolar electrodes on top and bottom surfaces. The directions of the polarization vector are noted by the block arrows. Length is 22.4mm, thickness is 0.7mm, and width is 3 mm (not shown). Displacements $u_z$ and $u_x$ occur at the tip. Panel (1) presents a standard single crystal cantilever, panel (2) presents two-inversely-poled domain cantilever, and panel (3) presents three-inversely-poled domain cantilever. All top electrodes marked by plus (+) are potential ones, and all bottom electrodes marked by minus (-) are grounded.

The acoustical displacements are calculated by a Finite Element code that was developed following the approach previously used to calculate dispersion curves for acoustic modes in multi-domain ferroelectric plates[11, 12]. The FEM routine is based on the Hamilton variational principle, where the solution is found by minimizing the time variation of the total energy that consists of four parts: $E_{kin} = 0.5\int \dot{u}^2 \, dx\, dz$ is the kinetic energy; $E_d = 0.5\int \mathbf{E} \cdot \mathbf{D} dx dz$ is the energy associated with electric field $\mathbf{E}$ and electrical displacement $\mathbf{D}$; $E_{st} = 0.5\int \mathbf{S}^{tr} \cdot \mathbf{T} dx dz$ is the elastic energy due to stress $\mathbf{T}$ and strain $\mathbf{S}$, where superscript *tr* denotes a transposed matrix; $W = \int \varphi q_s dx$ is work done by the excitation source when the electric potential $\varphi = V_0 \cos(2\pi f t)$ is applied to the electrodes; and $q_s$ is the surface charge density on the electrodes. The components of displacement $u_m$ (m = x, y, z) satisfy the equations of motion[13]

$$\rho \ddot{u}_i = \frac{\partial T_{ij}}{\partial x_j} \quad (1)$$

and piezoelectric relations

$$\frac{\partial T_{ij}}{\partial x_j} = \frac{\partial}{\partial x_j}\left[ c_{ijkl}^E \frac{\partial u_k}{\partial x_l} + (-1)^{(1-\delta_{x,m})(n-1)} e_{mij} \frac{\partial \varphi}{\partial x_m} \right], \quad (2)$$

$$\frac{\partial D_m}{\partial x_m} = \frac{\partial}{\partial x_m}\left[ (-1)^{(1-\delta_{x,m})(n-1)} e_{mij} \frac{\partial u_i}{\partial x_j} + \varepsilon_{mj}^S \frac{\partial \varphi}{\partial x_j} \right], \quad (3)$$

where: $c^E$ is the stiffness tensor under constant electric field, $e$ is the piezoelectric coupling tensor, $\varepsilon^S$ is the dielectric permittivity tensor under constant strain, and $\delta_{x,m}$ is the Kronecker symbol, $n$ is domain number 1, 2, and 3. The sign of piezoelectric coupling constants is inverted in the second domain ($n = 2$) for the Y and Z directions, but remains the same for the X direction. This represents a 180° rotation of the crystallographic coordinate system about the X axis in the inverted domains. The loss may be introduced by adding an imaginary part $c_{ijkl}''^{E}$ to the stiffness constants $c_{ijkl}^{E} = c_{ijkl}'^{E} + i c_{ijkl}''^{E}$, where $c_{ijkl}''^{E} = \delta c_{ijkl}'^{E}$, factor δ is some small parameter, and $c_{ijkl}'^{E}$ are $\varepsilon_{11}^S$=1.063x10$^{-8}$ C$^2$/Nm$^2$, $\varepsilon_{33}^S$=1.328x10$^{-8}$ C$^2$/Nm$^2$, $e_{31}$=-10.169 C/m$^2$,

$e_{33} = 18.662$ C/m², $e_{15} = 15.269$ C/m², $c_{11}^E = 21.130 \times 10^{10}$ N/m², $c_{12}^E = 14.430 \times 10^{10}$ N/m², $c_{13}^E = 15.997 \times 10^{10}$ N/m², $c_{33}^E = 21.409 \times 10^{10}$ N/m², $c_{44}^E = 3.215 \times 10^{10}$ N/m², $c_{66}^E = 1.585 \times 10^{10}$ N/m² used in the calculations. The attenuation parameter δ is related to mechanical quality factor $Q = 1/(2\delta)$. The equations (1) – (3) are solved together with the boundary conditions of electric potential amplitude $V_0$ at the top electrode, zero electric potential at the bottom electrode, and zero acoustical displacement at the left-side edge where the cantilever is clamped. In the FEM routine, a triangular numerical mesh with linear approximating functions is used. The solution is tested for convergence by gradually increasing mesh density until the results differed by less than 1%. The results of FEM-computations for the 1st, 2nd, and 3rd vibrational modes/resonances are given in Fig. 2 for three possible models including single domain (a), 2-domain (b) and 3-domain (c) cantilever.

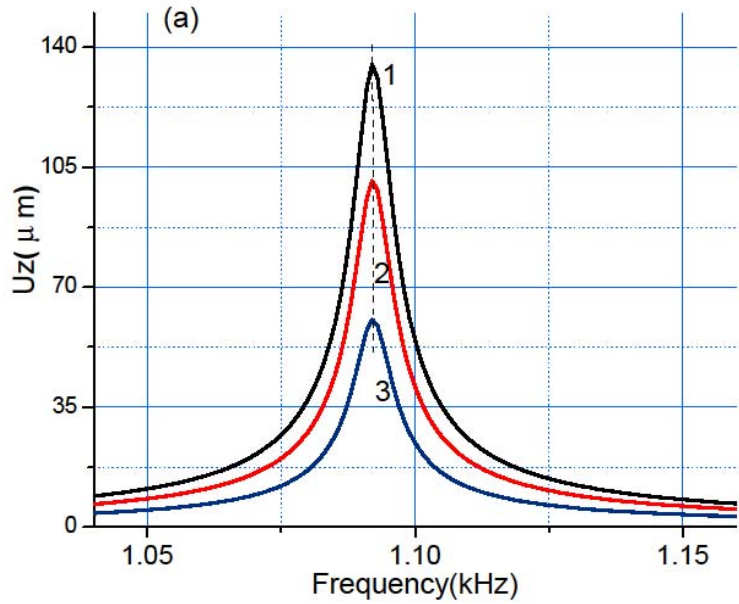

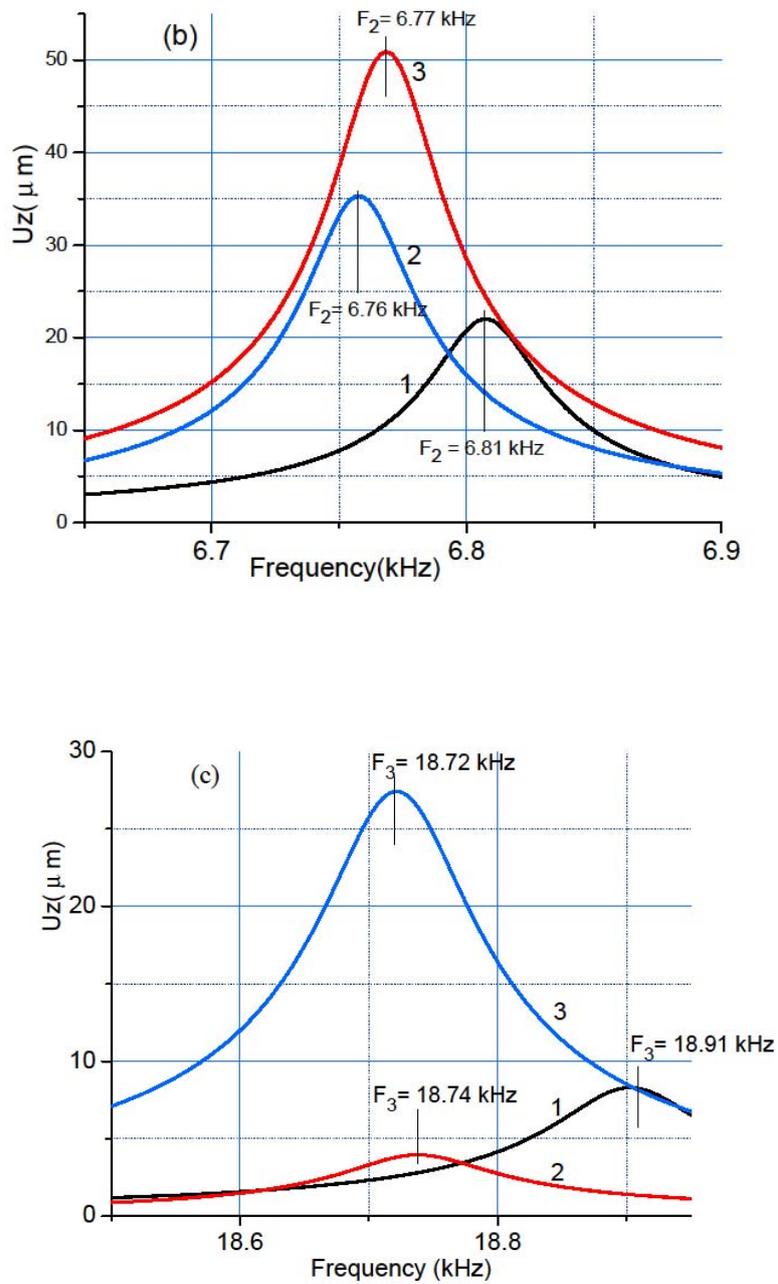

Fig. 2. Vertical displacement $u_z$ versus frequency in the multidomain piezoelectric cantilevers shown in Fig. 1, which are made of piezo-ceramic PIC-181 with Q = 100. Excitation by an electric field E = 14.3 kV/mm. Plots 1, 2 and 3 correspond to the single-domain, 2-domain, and 3-domain cantilever, respectively, as shown in Fig. 1, panels (1), (2) and (3). Panel (a) presents displacement amplitudes for the 1st vibrational mode at 1.09 kHz, panel (b) presents displacement amplitudes for the 2nd vibrational mode at 6.77 kHz, and panel (c) presents displacement amplitude for the 3rd vibrational mode at 18.72 kHz.

The second type of cantilever configurations may be designed with the help of the bipolar electrodes deposited on a single crystal prepolarized bar, as shown in Fig. 3. Actually, the first panel (1) is the same as in Fig. 1-(1), but the panels (2) and (3) are very different.

(1)

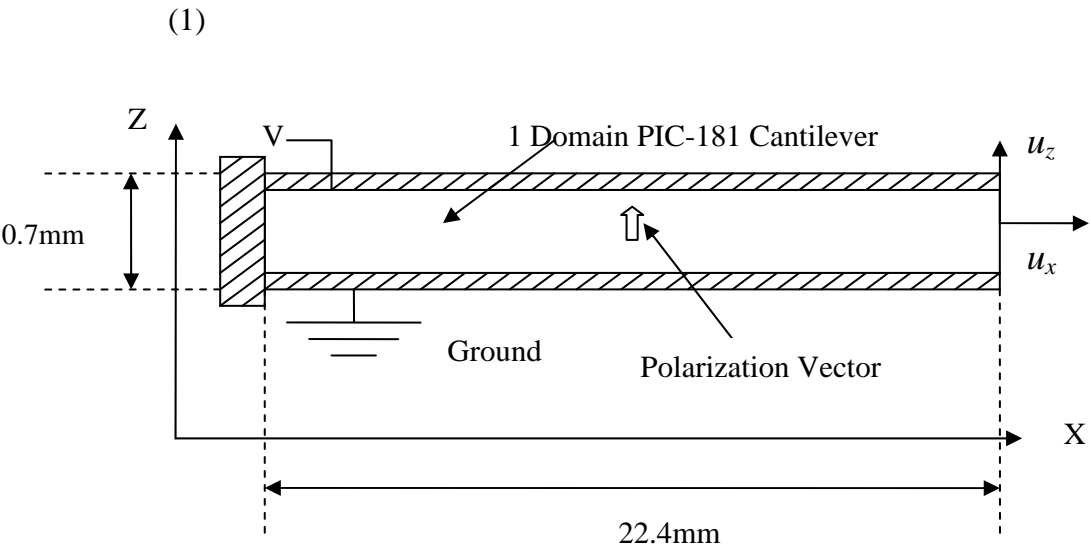

(2)

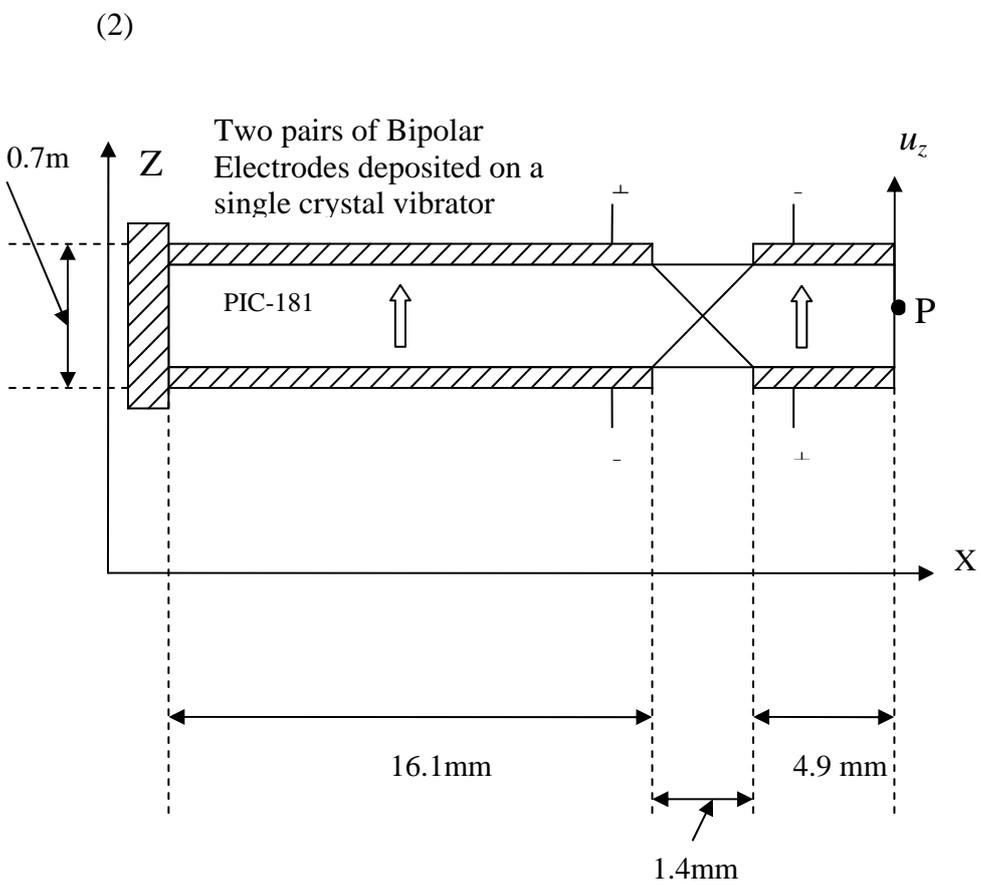

(3)

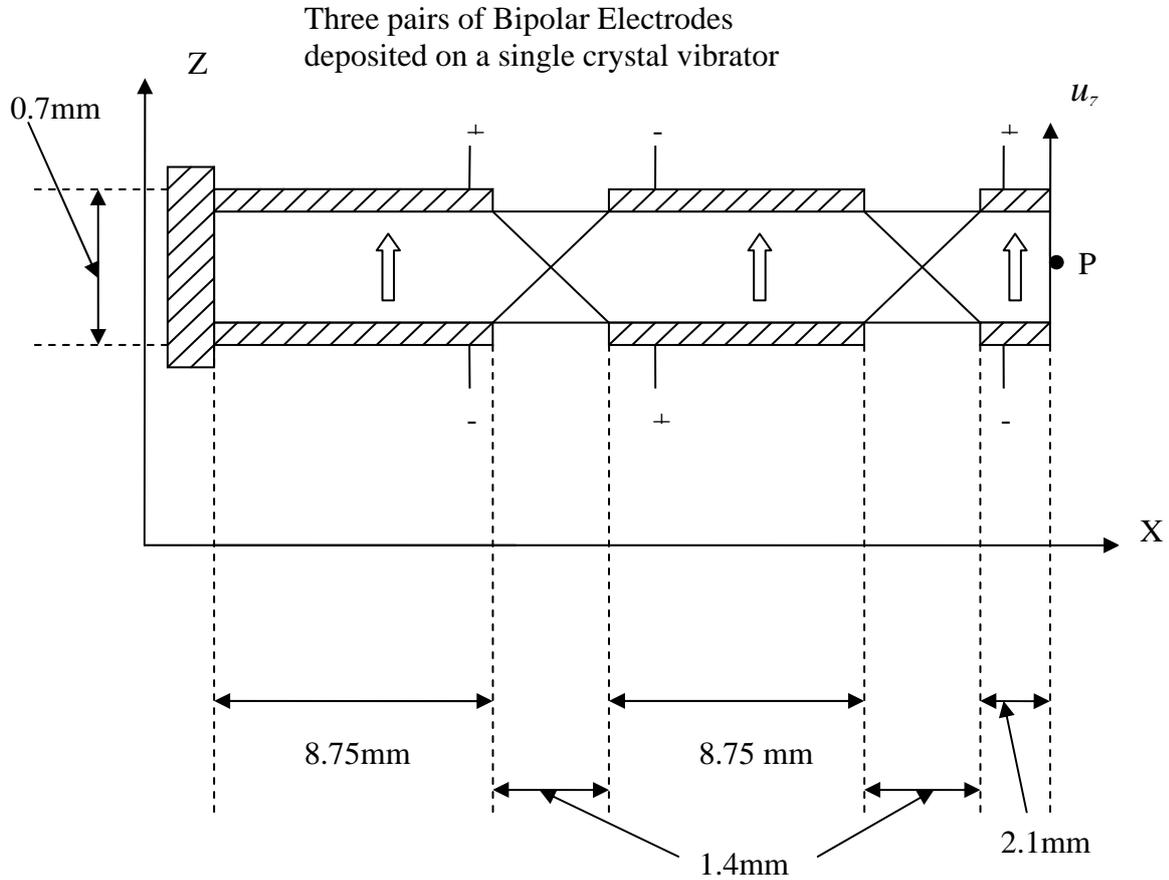

Fig. 3. The single crystal cantilevers with bypolar electrodes on top and bottom surfaces. The polarization vectors are along the z-axis. Length is 22.4mm, thickness is 0.7mm, and width is 3 mm (not shown). Displacements $u_z$ and $u_x$ occur at the tip. Panel (1) presents a standard single crystal cantilever similar to those in Fig. 1-(1), panel (2) presents two pairs of metal electrodes with positive and negative electric potential on both top and bottom surfaces, and panel (3) presents three pairs of metal electrodes with positive and negative electric potential on both top and bottom surfaces.

In the case of two-electrode bipolar excitation scheme, each electrode has a 1.4 mm gap at the domain boundary. The bottom electrode below the longer domain is connected to the top electrode above the shorter domain, and the top electrode above the longer domain is connected to the bottom electrode below the shorter domain. When this design is used, the inverse voltage is applied to the smaller domain when compared to the longer domain. Letter **P** in Fig.3 indicates the "probe" point, where the displacement component $u_z$ is calculated.

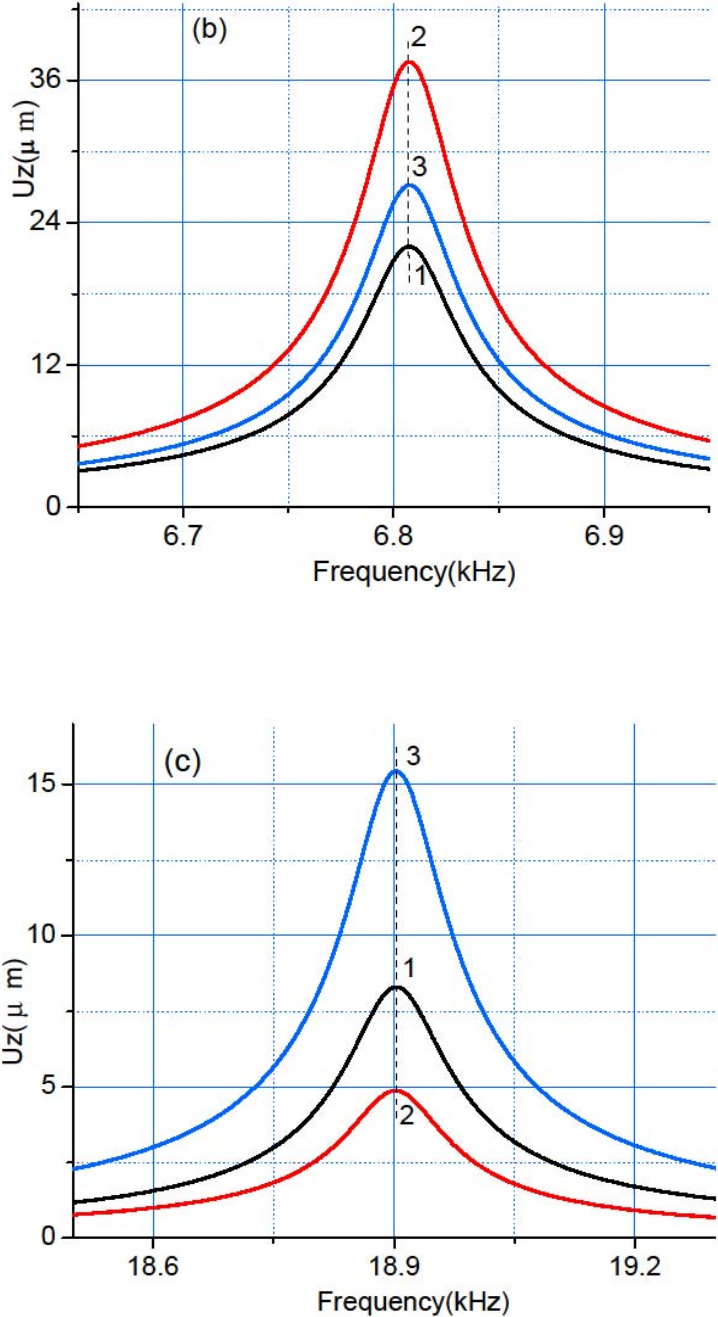

Fig. 4. Transversal vertical displacement $u_z$ versus frequency in the single crystal piezoelectric cantilevers with bipolar electrodes shown in Fig. 3-(2), (3). Piezo-ceramic PIC-181, Q = 100. Excitation by an electric field E = 14.3 kV/mm. Plot 1

corresponds to a single pair of electrodes as shown in Fig. 1-(1). Plots 2 and 3 correspond to the two and three pairs of electrodes, respectively, as shown in Fig. 3-(2) and Fig. 3-(3). Panel (b) presents displacement amplitudes for the 2$^{nd}$ vibrational mode at 6.81 kHz, panel (c) presents displacement amplitude for the 3$^{rd}$ vibrational mode at 18.9 kHz.

The FEM computation with the configuration shown in Fig. 3-(1) is the same as panel (a) of Fig. 2. However, the calculations for other configurations, Fig. 3-(2) and Fig.3-(3) are different from those in Fig. 2-(b) and Fig. 2-(c). These results are given in Fig. 4, panels (b) and (c), respectively.

At the first resonance, the only vibration node in the plate is at $x = 0$, where the cantilever is clamped. In the single domain plate, the $u_z$ component of displacement is coherent along the X direction. Two different settings can be used for excitation of the 2$^{nd}$ mode. The unipolar, when rf voltage is applied to top electrodes, and bottom contacts are grounded, like in Fig. 1-(2), and bipolar when rf voltage is applied to the top electrode with the bottom contact grounded in the 1$^{st}$ domain, and the electrode configuration is flipped in the 2$^{nd}$ domain, like in Fig. 3-(2).

The vibration amplitude $u_z$ for Q=100 at 2$^{nd}$ mode, or second resonance, is enhanced for the 2-domain cantilever contradictory to the single domain plate. The corresponding resonance frequency for the 2-domain plate is 6.8 kHz. The 2-domain cantilever shows an increase of the vibration amplitude compared to the single domain plate. The amplitude enhancement can be explained by comparing the displacement components at different positions $x$ along the cantilever length. The vibration component has a node at $x = 17.5$ mm. In the 2-domain plate, the domain boundary is placed at this node deliberately to improve the performance of the 2-domain cantilever compared to its single-domain counterpart. The vibration amplitude $u_z$ at $x = 22.4$ mm increases in the 2-domain cantilever in comparison to the single domain cantilever.

Another new configuration with three inversely poled domains as in Fig. 1-(3) also gives amplitude increase at higher, $3^{rd}$, vibrational mode, as clear from comparison of Fig. 2-(c) and Fig. 4-(c). It can be observed from these graphs that the amplitude $u_z$ for the 3-domain cantilever is greater than $u_z$ for the single domain cantilevers at $3^{rd}$ mode including the three pairs of bipolar electrode configuration, plot 3 in Fig. 4-(c).

**In conclusion:**

1) We propose a multi-domain cantilever model that has improved performance at selected resonance frequencies. 2) The vibration amplitude in the multi-domain cantilever with inversely poled 2 or 3 domains increases at the $2^{nd}$ or $3^{rd}$ resonance, respectively, if the domain boundaries are located at the nodes of vibration displacements. 3) Based on our calculations for several Q in the range from 10 to 100, the two-domain cantilever shows an advantageous increase in vibration amplitude at $2^{nd}$ harmonic resonance when compared to the single domain plate. 4) There is an advantage to multidomain cantilevers in that the number of domains can be optimally selected at appropriate resonance frequencies without changing the size of the device. 5) The applications for these results include microelectromechanical devices, where size may be an important parameter.

This work, in part, is made possible due to the research grant "Nonlinear vibrations of piezoelectric resonators," UM, 2011-2-12. We also are grateful to Dr. Lucien Cremaldi and Dr. Andriy Nadtochiy for useful discussions.

**References.**

[1]S. Banerjee, N. Gayathri, S. Dash, A. K. Tyagi, and B. Raj, "A comparative study of contact resonance imaging using atomic force microscopy", Appl. Phys. Lett. **86**, 211913 (2005).

[2]Ethem Erkan Aktakka, Rebecca L. Peterson, and Khalil Najafi. "High Stroke and High Deflection Bulk-PZT Diaphragm and Cantilever Micro Actuators and Effect of Pre-Stress on Device Performance." IEEE Jour. of Microelectromechanical Systems. 14 pgs. (2013) http://sitemaker.umich.edu/aktakka/files/aktakka_2013_jmems.pdf

[3]A. B. Nadtochiy, T. K. Hollis, and I. V. Ostrovskii, "Ferroelectric bimorph cantilever with self-assembled silane layer", Appl. Phys. Lett. **93**, 263503 (2008).

[4]R. G. Polcawich, J. S. Pulskamp, S. Bedair, G. Smith, R. Kaul, C. Kroninger, E. Wetzel, H. Chandrahalim, and S. A. Bhave, "Integrated PiezoMEMS actuators and sensors", IEEE Sensors, 2193 (2010).

[5]M. Sitti, D. Campolo', J. Yan, R. S. Fearing, T. Su, D. Taylor, and T. D. Sands, "Development of PZT and PZN-PT Based Unimorph Actuators for Micromechanical Flapping Mechanisms", Proceedings of the IEEE Int. Conf. on Robotics and Automation, 3839 (2001).

[6]Z. Cao, J. Zhang, H., Kuwano, IEEE Int. Conf. on Nano/Micro Engineering and Molecular Systems, 716 (2001).

[7]C. K. Lee and F. C. Moon, "Modal sensors/actuators". Journal of Applied Mechanics **57**(6), p. 434-441 (1990).

[8]Woo-Seok Hwang and Hyun Chul Park, "Finite element modeling of piezoelectric sensors and actuators", AIAA Journal **31**(5), p. 930 (1993).

[9]Jung-Kyu Ryou, Keun-Young Park, and Seung-Jo Kim, "Electrode Pattern Design of Piezoelectric Sensors and Actuators Using Genetic Algorithms", AIAA Journal **36**(2), p. 227-223 (1998).

[10] L. Pintilie and M. Lisca, "Polarization reversal and capacitance-voltage characteristic of epitaxial Pb(Zr,Ti)$O_3$ layers", Appl. Phys. Lett., **86**, 192902 (2005).

[11] I. V. Ostrovskii, V. A. Klymko, and A. B. Nadtochiy, "Plate wave stop-bands in periodically poled lithium niobate", JASA Express Lett. **125**(4), EL129 (2009).

[12] I. V. Ostrovskii, A. B. Nadtochiy, and V. A. Klymko, "Velocity dispersion of plate acoustic waves in a multidomain phononic superlattice", Phys. Rev. B **82**(1), 014302 (2010).

[13] B. A. Auld, "Acoustic fields and waves in solids", (R.E. Krieger Publishing Co., Inc.; Malabar, FL 32950, 1990), Vol. 2.